\documentclass[10pt,aps,prd,letterpaper,typeset,twocolumn,superscriptaddress,showpacs,floatfix]{revtex4-2}
\usepackage{color}
\usepackage{graphicx}
\usepackage{amsmath}
\usepackage{amssymb,amsthm}
\usepackage{hyperref}
\usepackage{mathtools}
\usepackage{physics}
\usepackage{multirow}
\usepackage{caption, subcaption}
 
\graphicspath{{pict/}{}}
\usepackage[utf8]{inputenc}
\usepackage{pgf,tikz}
\usepackage{mathrsfs}
\usepackage{bm}
\usepackage{array}
\usepackage{blkarray}
\usepackage{xcolor}
\usepackage{float}
\hypersetup{colorlinks=true,linkcolor=blue,citecolor=blue,urlcolor=blue}

\begin{document}

\title{Discrimination of metric theories}

\author{F. J. Lobo}
\affiliation{Facultad de Ciencias F\'isicas y Matem\'aticas, Departamento de F\'isica, Universidad de Concepci\'on, Concepci\'on, Chile}
\affiliation{Instituto Milenio de Investigaci\'on en \'Optica, Universidad de Concepci\'on, Concepci\'on, Chile}
\author{M. Rivera-Tapia}
\affiliation{Facultad de Ciencias F\'isicas y Matem\'aticas, Departamento de F\'isica, Universidad de Concepci\'on, Concepci\'on, Chile}
\affiliation{Instituto Milenio de Investigaci\'on en \'Optica, Universidad de Concepci\'on, Concepci\'on, Chile}
\author{G. Rubilar}
\affiliation{Facultad de Ciencias F\'isicas y Matem\'aticas, Departamento de F\'isica, Universidad de Concepci\'on, Concepci\'on, Chile}
\author{O. Jim\'enez}
\affiliation{Centro Multidisciplinario de F\'isica, Vicerrector\'ia de Investigaci\'on, Universidad Mayor, 8580745 Santiago, Chile}
\author{A. Delgado}
\affiliation{Facultad de Ciencias F\'isicas y Matem\'aticas, Departamento de F\'isica, Universidad de Concepci\'on, Concepci\'on, Chile}
\affiliation{Instituto Milenio de Investigaci\'on en \'Optica, Universidad de Concepci\'on, Concepci\'on, Chile}

\begin{abstract}
We study the possibility of discriminating between metric theories within the Parametrized Post-Newtonian formalism. In this approach, the two-dimensional quantum state of a massive quantum clock becomes, after propagating at low speed and in a weak gravitational field, a function of the post-Newtonian parameters and thus a signature of a metric theory. To discriminate among metric theories, we resort to quantum-state discrimination strategies such as minimum-error and unambiguous discrimination. In particular, we show that it is possible to refute the hypothesis that a particular metric theory describes spacetime with a single detection event and that it is possible to discriminate with certainty between two different metrics, also with a single detection event. In general, the success probability of the discrimination strategy is a harmonic function of the product of the difference of the proper time corresponding to each quantum clock state, the energy difference between the energy eigenstates of the quantum clock, the propagation length, and speed. It is thus possible to find suitable length and speed scales such that the success probability is close to one by selecting a quantum system with the highest energy difference and the longest natural lifetime. According to this, atomic nuclei such as thorium are considered the most suitable quantum clocks.  We also show that the use of a ensemble of quantum clocks leads to a significant increase in the probability of success in discriminating between post-Newtonian parameters that differ by $10^{-5}$. This facilitates achieving a probability of success approaching unity with distances on the scale of several kilometers and velocities approximating one-thousandth of the speed of light for a ensemble of only 10 quantum clocks.
\end{abstract}

\maketitle

\section{Introduction}
The Einstein Equivalence Principle (EEP) lays at the foundations of General Relativity (GR), a theory whose predictions have been consistently tested through challenging experiments and observations. Interestingly, besides GR, there are other theories that follow the EEP. This includes metric theories such as GR, where matter and non-gravitational fields respond only to spacetime, but consider additional fields besides the metric, such as scalar fields, vector fields, and so on. These additional fields modify the manner in which matter and non-gravitational fields generate gravity and determine the metric \cite{Clifton2012}. Once the metric has been determined, it is the only object that acts on the matter as prescribed by the EEP. Thus, metric theories are distinguished from each other by the particular way in which matter and possibly other gravitational fields determine the metric \cite{Shankaranarayanan2022, Papantonopoulos2015}.
The existence of alternative metric theories consistent with the EEP and competing with GR has led to the study, comparison, and experimental discrimination between them. A preferred approach to accomplish this is the use of the parametrized post-Newtonian (PPN) formalism \cite{WillReview}. Within this approach, which holds in the slow velocity and weak gravitational field limit, the spacetime metric is perturbatively expanded about the Minkowski metric plus various terms depending on gravitational potentials and a set of dimensionless parameters. The values of these parameters distinguish a particular metric theory \cite{Hohmann2017,PhysRevLett.99.241101, PhysRevD.105.044002}.

 The value of the post-Newtonian parameters can be obtained or constrained by experiments. These are based on the direct measurement of quantities that are functions of post-Newtonian parameters. For this purpose, the deflection of light and its time delay \cite{Will_2018,Shapiro2004,Bertotti2003}, as well as the perihelion shift of Mercury \cite{Will_2018,Verma2014}, have been used. Recently, it has been considered \cite{Saleem2022} that the improved sensitivity \cite{Abbott2018} of the Advanced LIGO \cite{Aasi_2015}, Advanced Virgo \cite{Acernese_2015}, and KAGRA \cite{Somiya_2012} detectors could lead to high-precision tests of GR deviations using gravitational wave detection.

Instead of measuring or constraining the values of the post-Newtonian parameters, here we study the possibility of distinguishing between metric theories within the PPN formalism using quantum hypothesis testing. In classical hypothesis testing \cite{CaseBerg:01}, an agent attempts to determine with the least possible error whether the available information preferentially supports a particular hypothesis among a set of hypotheses. The available information is modeled as a sample from a known probability distribution defined by a fixed set of parameters, while each hypothesis is defined by a region in the parameter space. In quantum hypothesis testing \cite{Cheng_2025}, the available information corresponds to the results of a quantum measurement, which is designed to minimize a predefined error metric, while each hypothesis corresponds to a quantum state. In this scenario, a single measurement result could provide enough information to rule out a given hypothesis.

We consider the propagation of a massive quantum clock, that is, a massive particle endowed with a two-dimensional inner degree of freedom, at low speed and in a weak gravitational field. The final quantum state of the clock is a function of the post-Newtonian parameters that characterize a metric theory and plays the role of a hypothesis. The quantum clock states of various metric theories are mutually non-orthogonal and, consequently, cannot be discriminated deterministically and with certainty at the same time.  Thereby, as a quantum measurement, we use various strategies for discriminating non-orthogonal quantum states \cite{Barnett:09,Bergou10022010}. These strategies are defined by quantum measurements designed to identify, with a single measurement result, quantum states generated from a known list of quantum states with a known probability distribution.
Quantum state discrimination is a fundamental task in quantum information processing, essential to extract classical information from quantum systems \cite{Bae2015}. It plays a crucial role in various applications, including quantum communication, entanglement, cryptography, and computing, and is deeply connected to fundamental principles such as relativistic causality and the no-signaling constraint. Quantum state discrimination strategies have already been demonstrated experimentally \cite{Barnett01061997,Clarke2001,Cook2007,Wittmann2008}, even in higher-dimensional quantum systems \cite{Mohseni2004,Waldherr2012,Becerra2013,Agnew2014,Delgado2017,Solis-Prosser_2022}, achieving excellent agreement with theoretical predictions. We first show a simple scheme, based on a von Neumann measurement, that allows for ruling out the hypothesis that a particular metric theory describes spacetime. In the case of two competing metric theories, we use minimum-error discrimination. This makes it possible to determine which of the metric theories is verified, although with a minimal probability of error. This result can be improved using unambiguous state discrimination, which allows to distinguish between two metric theories without misidentification error, albeit resorting to a generalized quantum measurement and with a smaller success probability than using minimum-error discrimination. 

The success probability of each one of these three approaches is a harmonic function of the product of the difference of the proper time corresponding to each quantum clock state, the energy difference between the energy eigenstates of the quantum clock, the propagation length and speed. Typically, this probability can be increased by considering a large propagation length.  However, using a quantum internal degree of freedom as a clock requires maintaining the coherence of the corresponding energy levels throughout the clock's propagation. Therefore, quantum systems whose energy eigenstates have a long natural lifetime are favored. Thus, the most promising candidates for quantum clocks are low-energy states of bare atomic nuclei such as Th-229 \cite{Beeks2021,Yamaguchi2024,Tiedau2024}.  The radiative lifetime of the 8.4 eV nuclear isomer state in Th-229, when embedded in a CaF$_2$ crystal, is close to 630 s \cite{Tiedau2024,Kraemer2023}. Theoretical studies suggest that this can be between $10^3$ and $10^4$ s in vacuum \cite{Tkalya2015,Minkov2017}. 

This nuclear isomer has been resonantly excited \cite{Tiedau2024} using a tunable benchtop laser system, which enables coherent manipulation of the transition between the ground and nuclear isomer states. Other choices such as atoms, molecules, electrons, and neutrons lead to vanishingly small discrimination probabilities. A challenging scenario occurs when discriminating metric theories that have very close post-Newtonian parameters, for example, when trying to rule out a metric theory whose parameters are within the uncertainty of a metric that we suspect describes spacetime. In this case, propagation lengths on the order of $10^5$ km and speeds on the order of $10^3$ km/s allow us to reach success probabilities of the order of $10^{-2}$.   Similar probability values can be achieved considering shorter lengths and higher speeds, for example, a propagation length of $3.2$ km and a speed of $300$ km/s lead to a success probability of $7\times10^{-2}$. Higher probabilities can be achieved on larger length or speed scales. Additionally, we also show that using an ensemble of identically and independently prepared quantum clocks also increases the probability of success. This generally exhibits an exponential increase towards 1 as the size of the ensemble is augmented. Under certain conditions, an ensemble of just 10 quantum clocks leads to a success probability close to 0.999.

This article is organized as follows. In Section \ref{SECTION2} we briefly review the post-Newtonian formalism, the method used to couple gravity to a quantum system, and quantum state discrimination strategies. In Section \ref{SECTION3} we present our main results on the discrimination of metric theories, and in \ref{SECTION4} we present our conclusions.

\section{A brief review of the PPN formalism, the coupling of quantum systems to gravity, and the discrimination of quantum states.}
\label{SECTION2}

In this section, we briefly review the post-Newtonian parameterization of a spacetime metric and introduce a Hamiltonian operator for a low-speed massive particle interacting with a weak gravitational field. We then review the problem of quantum-state discrimination and related strategies.

\subsection{Parametrized post-Newtonian formalism}

The Parametrized Post-Newtonian (PPN) formalism provides a framework for comparing metric theories of gravity by quantifying deviations from Newtonian gravity in the weak-field, slow-motion limit. This formalism utilizes a set of parameters, each with a distinct physical interpretation \cite{1972ApJ...177..757W, WillReview}: $\gamma$ determines the amount of space curvature produced by mass; $\beta$ characterizes the non-linearity of gravity, reflecting the extent to which gravity itself generates gravity. The PPN formalism also includes parameters that assess the effects of a preferred universal rest frame on the post-Newtonian metric, which are relevant to theories that propose such a frame. These parameters quantify how deviations from this hypothetical universal rest frame affect gravitational interactions. In addition, the formalism incorporates parameters that quantify potential violations of energy-momentum conservation laws. These parameters would be non-zero if, in certain theoretical frameworks, the fundamental principles of energy and momentum conservation were not strictly adhered to. These parameters, particularly $\gamma$ and $\beta$, are crucial to analyze experimental tests of gravity, such as the classical tests of General Relativity. In this work, we use a simple version of the parameterized post-Newtonian approximation, which depends only on the parameters $\gamma$ and $\beta$. In this case, the parameterized spacetime metric is given by
\begin{eqnarray}
    g_{00} &=& - \left[1+2\frac{\phi}{c^{2}} + 2 \beta \left(\frac{\phi}{c^{2}}\right)^{2}\right],\nonumber\\
    g_{0i} &=&0,\nonumber\\
    g_{ij} &=& \left(1- 2\gamma \frac{\phi}{c^{2}}\right)\delta_{ij},
    \label{PPN_metric}
\end{eqnarray}
where the coefficient $g_{00}$ is approximated up to second order in $\phi/c^2$, where $\phi$ is the Newtonian potential 
\begin{eqnarray}
    \phi(\vec{x},t)=-G\int \frac{\rho(\vec{x}',t)}{|\vec{x}-\vec{x}'|} d^{3}\vec{x}',
\end{eqnarray}
and $\rho$ is the rest mass density of matter generating the gravitational field. The inverse of the metric is
\begin{eqnarray}
    g^{00}&=& - \left[1-2\left(\frac{\phi}{c^{2}}\right) +2(2-\beta)\left(\frac{\phi}{c^{2}}\right)^{2}\right],\nonumber\\
    g^{ij}&=& \left[ 1+2\gamma\left(\frac{\phi}{c^{2}}\right) + 4\gamma^{2}\left(\frac{\phi}{c^{2}}\right)^{2}\right]\delta^{ij},
    \label{PPN_metric_inverse}
\end{eqnarray}
which holds up to second order in $\phi/c^2$ and where we use the mostly plus convention for the spacetime metric $g^{\mu \nu}$ Eq.~\eqref{PPN_metric_inverse}.

Einstein's general relativity is characterized by $\gamma = \beta=1$. Our goal is to distinguish between several gravitational theories through their dependence on $\gamma$ and $\beta$.

\subsection{Hamiltonian operator for a massive particle in a gravitational field}

To derive a Hamiltonian operator for the interaction between the center of mass of a quantum clock, that is, a massive quantum system with an inner degree of freedom, with a gravitational field \cite{Zych2011, zych2017quantum, Paczos2024quantumtimedilation}, we begin with the solution $\varphi_{KG}$ of the Klein-Gordon equation:
\begin{equation}
(\hbar^{2}g^{\mu \nu}\nabla_{\mu} \nabla_{\nu}+ M^{2}c^{2})\varphi_{KG} = 0,
\end{equation}
for a quantum clock with rest mass $M$. Assuming the ansatz $\varphi_{\rm KG} = e^{iS(x,t)/\hbar}$, we expand the phase $S(x,t)$ in powers of the speed of light $c$ in the weak-field limit, that is, $S(x,t)=c^{2}S_0(x,t) + S_1(x,t) + c^{-2}S_2(x,t) + O(c^{-4})$. This leads to a modified Schr\"odinger equation given by the Hamiltonian operator \cite{PhysRevA.100.052116, LAMMERZAHL199512, PhysRevD.44.1067}
\begin{eqnarray}
    H &=&  Mc^{2} + \frac{p^{2}}{2M} - \frac{p^{4}}{8M^{3}c^{2}} +M\phi(r) 
    \nonumber\\
    &+&  \frac{\left(2\gamma + 1 \right)}{2Mc^{2}} \left( \phi(r)p^{2} + \left[ \vec{p}\, \phi\right] \cdot \vec{p}\right) \nonumber\\
    &-&  \left(\frac{1}{2} - \beta\right)M \frac{\phi^{2}(r)}{c^{2}} + \frac{3}{4} \frac{\gamma}{Mc^{2}}\left[ p^{2} \phi(r)\right],
    \label{Modified_Hamiltonian_PPN}
\end{eqnarray}
where $\vec{p}$ is the momentum operator, $\phi(r)$ the gravitational potential, and the square brackets $[\;]$ denote the action of $\vec{p}$ on $\phi$. This Hamiltonian operator acts on the Hilbert space $\mathcal{H}_{\text{cm}}$ that describes the center of mass of the quantum clock.

In addition to $\mathcal{H}_{\text{cm}}$, we consider a two-dimensional inner degree of freedom with the Hilbert space $\mathcal{H}_{\text{clock}}$ that can be physically implemented, for example, as the electronic energy levels of an atom or the vibrational energy levels of a molecule.  This internal degree of freedom plays the role of a clock \cite{zych2017quantum}. The total Hilbert space is thus $\mathcal{H}_{\text{tot}}=\mathcal{H}_{\text{cm}}\otimes\mathcal{H}_{\text{clock}}$. The Hamiltonian operator in the rest frame driving the evolution of the inner degree of freedom is
\begin{equation}\label{H_clock}
H_{\rm clock} = E_0 | 0 \rangle \langle 0 |  + E_1 | 1 \rangle \langle 1 |,
\end{equation}
where $|0\rangle$ and $|1\rangle$ are eigenstates with energies $E_0$ and $E_1$, respectively. 

The total energy of the system at rest not only includes the energy of the rest mass $M$ but also the energy of the internal degree of freedom of the particle $H_{\rm clock}$. Considering the total mass $M$ of the system at rest, we have
\begin{equation}
    M = m + \frac{H_{\rm clock}}{c^{2}},
    \label{Mass-Energy_Quantum}
\end{equation}
where $m$ is the mass of the system for the state in which $\langle H_{\rm clock}\rangle=0$. 

Therefore, to include the inner degree of freedom in the Hamiltonian Eq.~\eqref{Modified_Hamiltonian_PPN}, we use  Eq.\thinspace\eqref{Mass-Energy_Quantum} and approximate up to first order in $H_{\rm clock}/(mc^2)$, which leads to the Hamiltonian operator
\begin{eqnarray}
    H = H_{\rm cm}^{\rm rel}\otimes\mathbb{I} & + \lambda\otimes H_{\rm clock},
\label{TotalHamiltonian}
\end{eqnarray}
where $H_{\rm cm}^{\rm rel}$ is the Hamiltonian operator of the center of mass of the quantum clock given by
\begin{eqnarray}
    H_{\rm cm}^{\rm rel} &=&  mc^{2} + \frac{p^{2}}{2m} - \frac{p^{4}}{8m^{3}c^{2}} +m\phi(r)
    \nonumber\\
    &+&  \frac{\left(2\gamma + 1 \right)}{2mc^{2}} \left( \phi(r)p^{2} + \left[ \vec{p}\, \phi\right] \cdot \vec{p}\right) \nonumber\\
    &-&\left(\frac{1}{2} - \beta\right)m \frac{\phi^{2}(r)}{c^{2}} + \frac{3}{4} \frac{\gamma}{mc^{2}}\left[ p^{2} \phi(r)\right]
    \label{H_cm}
\end{eqnarray}
and the coupling of the gravitational field to the internal degree of freedom is given by the operator $\lambda$ defined as
\begin{eqnarray}
     \lambda &=& 1 - \frac{p^{2}}{2m^2 c^2} + \frac{3p^{4}}{8m^{4}c^{4}} +\frac{\phi(r)}{c^{2}}
     \nonumber\\
     &-&  \frac{\left(2\gamma + 1 \right)}{2m^2 c^{4}} \left( \phi(r)p^{2} + \left[ \vec{p}\, \phi\right] \cdot \vec{p}\right)  \nonumber \\ 
    &-&  \left(\frac{1}{2} - \beta\right)\frac{\phi^{2}(r)}{c^{4}} 
    - \frac{3}{4} \frac{\gamma}{m^2 c^{4}}\left[ p^{2} \phi(r)\right].
    \label{Coupling_operator}
\end{eqnarray}
As we can see from the previous expressions, the Hamiltonian operator $H_{\rm cm}^{\rm rel}$ depends linearly on $\gamma/c^2$ and $\beta/c^2$, while the operator $\lambda$ depends on $\gamma/c^4$ and $\beta/c^4$. 

\subsection{Quantum discrimination theory}

The problem of discrimination of quantum states consists in designing a quantum measurement to determine in which state a quantum system was prepared from a set $\{\rho_j\}$ of possible states generated with a priori probabilities $\{\eta_j\}$, respectively. The measurement is designed to satisfy a predetermined optimality criterion \cite{HELSTROM1967254,HELSTROM1968156,Helstrom1969QuantumDA}. Since quantum state discrimination can be formulated as a communication problem and involves optimal measurements, it finds wide application in quantum information processing and quantum foundations \cite{RevModPhys.74.145,Chefles01112000}.  

According to quantum theory, non-orthogonal quantum states cannot be deterministically distinguished and without error. Any quantum strategy will involve a non-zero probability of erroneous or inconclusive results \cite{Bae2015,Barnett:09,Bergou10022010}. Since a quantum-state discrimination strategy is based on the optimization of a cost function, it lacks, in general, analytic results. Typically, the optimization problem can be numerically solved using semi-definite programming \cite{Watrous2018} or stochastic optimization \cite{Concha2023}. Exceptions are cases of discrimination between only two quantum states or between sets of states with high symmetry, where the optimization problem is solved in a low-dimensional parameter space. In this scenario, we consider two discrimination strategies: minimum-error (ME) discrimination and optimal unambiguous discrimination (UD).

\subsubsection{Minimum-error discrimination}

The purpose of ME discrimination is to minimize the average probability of misidentifying the state of a quantum system \cite{barnett2009quantum,holevo,yuen}. This strategy involves finding the optimal Positive Operator-Valued Measure (POVM) $\{\Pi_i\}$ that minimizes the average probability of misidentifying the state \cite{qiu2008minimumerror}. If the measurement returns the result $i$, then we assume that the state prior to the measurement was $\rho_i$. The probability of this event is given by Born's rule $P(i|\rho_i)=Tr(\rho_i\Pi_i)$. An error, that is, a misidentification, occurs with probability $P(j|\rho_i)=Tr(\rho_j\Pi_i)$ when the state $\rho_i$ is identified as the state $\rho_j$ with $i\neq j.$ The average probability of error is thus $P_e(\{\Pi_i\})=1-\sum_i\eta_iTr(\rho_i\Pi_i)$. This is optimized on the set of POVMs to obtain the minimal average probability of error, that is,
\begin{equation}
P_H = \min_{\{\Pi_i\}} P_e(\{\Pi_i\}).
\end{equation}
The measurement that reaches minimum is known as the Helstrom measurement \cite{helstrom}, and the corresponding minimum average error probability is called the Helstrom bound \cite{barnett2009quantum}. 

In the case of discriminating between two states $\rho_1$ and $\rho_2$, with prior probabilities $\eta_1$ and $\eta_2$, respectively, the Helstrom bound has a simple analytical expression \cite{barnett2009quantum}
\begin{equation}
P_H= \frac{1}{2} \left( 1 - \mathrm{Tr} \left( \left| \eta_1 \rho_1 - \eta_2 \rho_2 \right| \right) \right),
\end{equation}
where $|A|=\sqrt{AA^\dagger}$. For two pure states $\rho_0=|\psi_0\rangle\langle\psi_0|$ and $\rho_1=|\psi_1\rangle\langle\psi_1|$, the previous expression can be further simplified to
\begin{equation}
P_H= \frac{1}{2} \left( 1 -\sqrt{1-4\eta_0\eta_1|\langle\psi_0|\psi_1\rangle|^2}\right)
\label{HelstromBoundpurestates}
\end{equation}
and the Helstrom measurement becomes a von Neumann, or projective, measurement  given by the observable $\Gamma=(1/2)(|\psi_0\rangle\langle\psi_0|-|\psi_1\rangle\langle\psi_1|)$. The eigenproyector $\Pi_0$ ($\Pi_1$) with positive (negative) eigenvalue identifies the state $|\psi_0\rangle$ ($|\psi_1\rangle$).

ME discrimination has been experimentally demonstrated using two-dimensional quantum systems \cite{Barnett01061997,Clarke2001,Waldherr2012} and higher-dimensional quantum systems \cite{Delgado2017}.

\subsubsection{Unambiguous discrimination}

An important feature of ME discrimination is that it minimizes the probability of a wrong identification. However, this is different from zero. UD arises as an attempt to perfectly distinguish quantum states. In general, this requires a POVM $\{\Pi_i\}$ ($i=0,1,\dots,n$) with a number of outcomes equal to $n+1$, where $n$ is the total number of states to be discriminated. The additional measurement outcome $i=0$ does not convey any information about the state to be identified, that is, it is not conclusive. The remaining measurement results $i=1,\dots,n$ allow for a perfect identification of each state, that is, they are unambiguous.

In the case of discriminating two non-orthogonal pure states $|\psi_0\rangle$ and $|\psi_1\rangle$ with a priori probabilities $\eta_0$ and $\eta_1$, respectively, the optimal average success probability $P_{UD}$, which considers the conclusive events only, depends on the relationship between the a priori probabilities and the inner product $s=|\langle\psi_0|\psi_1\rangle|$ \cite{JAEGER199583}. If $\eta_1<s^2/(1+s^2)$, then $P_{UD}=1-\eta_1-\eta_2s^2$. If $\eta_1>1/(1+s^2)$, then $P_{\rm{UD}}=1-\eta_2-\eta_1s^2$. If $s^2/(1+s^2)\leq\eta_1\leq1/(1+s^2)$, then $P_{UD}=1-2\sqrt{\eta_1\eta_2}s$. The particular case $\eta_1=\eta_2=1/2$ is known as the Ivanovic-Dieks-Peres limit \cite{IVANOVIC1987257,DIEKS1988303,PERES198819}, where 
\begin{equation}
P_{\rm {UD}}=1-s.    
\label{ProbabilityUD}
\end{equation}

\section{Discrimination between metric theories}
\label{SECTION3}

Our main objective is to distinguish between metric theories that obey the EEP. To do so, we assume that an observer, Alice, initially prepares the degree of freedom of motion of the center of mass of a massive quantum system into a well-localized Gaussian state. The internal degree of freedom, or quantum clock, is also prepared by Alice in an equally weighted superposition of energy eigenstates. The quantum system then propagates toward a second observer, Bob, along a geodesic curve only under the influence of the gravitational field, after which the quantum clock is described by a quantum state that depends on the post-Newtonian parameters $\gamma$ and $\beta$. This dependence allows us to distinguish between metric theories. The second observer performs measurements on the quantum system aimed at determining the particular metric theory that effectively generated the quantum clock state.

However, the possible final states of the quantum clock are mutually nonorthogonal, and consequently there is no quantum measurement that can discriminate between them with certainty and deterministically. This suggests the use of quantum-state discrimination strategies. In this section, we calculate the final state of the quantum clock as a function of post-Newtonian parameters and apply quantum-state discrimination strategies.

\subsection{Evolution of the state of the quantum clock}

We consider a quantum system that has one external and one internal degree of freedom. The initial separable state of the quantum system is
\begin{equation}
    \ket{\Psi(0)} = \ket{\psi_{0}}\otimes\ket{\tau},
    \label{Q_state_total}
\end{equation}
where $\ket{\psi_{0}}$ is the state of the center of mass, which is a solution of the free non-relativistic Schr{\"o}dinger equation governed by the Hamiltonian $H_0 = \hat{p}^2/(2m)$, and the state $\ket{\tau}$ describes the inner degree of freedom that plays the role of quantum clock. This is prepared in an equally weighted superposition of two energy eigenstates, $\ket{0}$ and $\ket{1}$ with energies $E_0$ and $E_1$, respectively, such that $\ket{\tau} = (\ket{0}+\ket{1})/\sqrt{2}$.

The system evolves to the final state (for further details see \cite{Zych2011, PhysRevD.55.1964, Rivera-Tapia_2022})
\begin{equation}
\ket{\Psi( t)}=e^{-\frac{i}{\hbar}\Delta t \langle H^{\mathrm{rel}}_{\mathrm{cm}} \rangle}\ket{\psi_0}
 \otimes \frac{e^{-\frac{i}{\hbar}\langle\bar{\lambda}\rangle E_0}}{\sqrt{2}} \left(\ket{0} + e^{-\frac{i}{\hbar}\langle\bar{\lambda}\rangle \Delta E}\ket{1}\right),
 \label{Q_state_final}
\end{equation}
where $\Delta E=E_1-E_0$, averages are taken with respect to the state $|\psi_0\rangle$, and the coupling has been redefined as $\langle\bar{\lambda}\rangle=\int dt \langle\lambda \rangle$ (see Appendix \ref{Appendix_Phase_shift} for details).

The previous expression shows that the external and the inner degrees of freedom are in a separable state. Since the measurements we will study do not involve the center-of-mass motion, this degree of freedom can be traced out while the quantum clock stays in the pure state
\begin{equation}
\ket{\psi}=\frac{1}{\sqrt{2}}(\ket{0} + e^{-\frac{i}{\hbar}\langle\bar{\lambda}\rangle \Delta E}\ket{1}),
\label{quantumclockstate}
\end{equation}
in which a global phase has been excluded. Thus, the phase that the moving clock acquires due to its interaction with the gravitational field enters the state $\ket{\psi}$ and depends on the post-Newtonian parameters $\gamma$ and $\beta$ through the function $\langle\bar{\lambda}\rangle$. Therefore, the final state of a quantum clock becomes a signature of a metric theory. For two of these states,
\begin{equation}
\ket{\psi_1}=\frac{1}{\sqrt{2}}(\ket{0} + e^{-\frac{i}{\hbar}\langle\bar{\lambda}_1\rangle \Delta E}\ket{1})
\end{equation}
and
\begin{equation}
\ket{\psi_2}=\frac{1}{\sqrt{2}}(\ket{0} + e^{-\frac{i}{\hbar}\langle\bar{\lambda}_2\rangle \Delta E}\ket{1}),
\end{equation}
the absolute value of inner product is

\begin{equation}
|\langle\psi_1|\psi_2\rangle|=|\cos({\frac{\Delta E}{2\hbar}(\langle\bar{\lambda}_1\rangle-\langle\bar{\lambda}_2\rangle)})|,    
\end{equation}
which in general does not vanish. The average $\langle\Delta\bar{\lambda}\rangle=\langle\bar{\lambda}_1\rangle-\langle\bar{\lambda}_2\rangle$ is evaluated with $|\psi_0\rangle$ a gaussian state. In this case, it can be shown that
\begin{equation}\label{eq:mean-lambda}
\langle\Delta\bar{\lambda}\rangle=\Delta\bar{\lambda}+\Delta\bar{\lambda}_{\rm {quant}},    
\end{equation}
where $\Delta\bar\lambda=\bar{\lambda}_1-\bar{\lambda}_2$ is the difference in arrival times given by 
\begin{equation}
    \Delta\bar\lambda = L\frac{\phi(R)}{c^{4}}\left( (\gamma_1-\gamma_2)v - (\beta_1-\beta_2)\frac{\phi(R)}{v}\right),
    \label{Difference_lambda}
\end{equation}
where we have considered that quantum particle propagates a distance $L$ at a constant speed $v$, so we can parameterize the total coordinate time as $t = L/v$. Also, Eq.~\eqref{Difference_lambda}
 is the difference in proper times between both metric theories. This quantity turns out to be independent of the mass $m$ of the quantum system. This expression has the same functional dependence as the operator $\lambda$ Eq.~(\ref{Coupling_operator}) but evaluated on the parameters of the gaussian state. The quantity $\Delta\bar{\lambda}_{\rm quant}$ is proportional to $\hbar$. Within the parameter values considered here, numerical simulations indicate that this term is extremely small and can be neglected. Thus, we replace $\langle\Delta\bar{\lambda}\rangle=\Delta\bar{\lambda}$ (see Appendix \ref{Appendix_Phase_shift} for details). In our numerical simulations, we have considered a localized Gaussian state for the center of mass. However, it is possible to consider states with even better localization properties \cite{PhysRevResearch.3.013049} that are similar to squeezed states.

\subsection{Simple discrimination scheme}

According to the results of the previous subsection, there are as many quantum clock final states as there are conceivable metric theories. This makes the problem of discrimination among metric theories especially difficult. However, a simple picture can be obtained if one tries to rule out or refute a given metric theory with $\gamma=\gamma_1$ and $\beta=\beta_1$ from any other metric theory.

For this purpose, we start by noting that the Hilbert space of the quantum clock is two-dimensional. Thus, if the state $\ket{\psi_1}$ corresponds to a metric theory with $\gamma_1$ and $\beta_1$ and the state $\ket{\psi_1^\perp}$ is orthogonal to $\ket{\psi_1}$, the two-dimensional identity operator can be represented as
\begin{equation}
\mathbb{I}=\ket{\psi_1}\bra{\psi_1}+\ket{\psi_1^\perp}\bra{\psi_1^\perp}.
\end{equation}
Therefore, we can cast the final quantum clock state $\ket{\psi_2}$ of any other metric theory with $\gamma=\gamma_2$ and $\beta=\beta_2$ as
\begin{equation}
\ket{\psi_2}=\bra{\psi_1}\psi_2\rangle\ket{\psi_1}+\bra{\psi_1^\perp}\psi_2\rangle\ket{\psi_1^\perp}.
\end{equation}
This suggests a measurement of the observable
\begin{equation}
O=\kappa\Pi + \kappa^{\perp}\Pi^\perp,
\label{ObservableO}
\end{equation}
with $\Pi=\ket{\psi_1}\bra{\psi_1}$ and $\Pi^\perp=\ket{\psi_1^\perp}\bra{\psi_1^\perp}$. 
Measurement of this observable leads to the detection of the eigenvalue $\kappa^\perp$ with probability $P(\kappa^\perp)=|\bra{\psi_1^\perp}\psi_2\rangle|^2=1-|\bra{\psi_1}\psi_2\rangle|^2$, which generally does not vanish. Thus, a single detection of $\kappa^\perp$ would lead to the conclusion that spacetime is described by a metric theory with $\gamma\neq\gamma_1$ or $\beta\neq\beta_1$. The detection of the eigenvalue $\kappa$ does not allow us to conclude that spacetime is described by a metric theory with $\gamma_1$ and $\beta_1$, since both states $|\psi_1\rangle$ and $|\psi_2\rangle$ have a component in state $|\psi_1\rangle$.

The probability of detecting the eigenvalue $\kappa^\perp$ is

\begin{equation}
    P(\kappa^\perp)=\sin^2\left(\frac{\Delta E}{2\hbar}\Delta\bar\lambda\right),
    \label{Prob_delta_first_proposal}
\end{equation}
 
Where $\Delta\bar\lambda=\bar{\lambda}_1-\bar{\lambda}_2$ s the difference in arrival times given by Eq.~\eqref{Difference_lambda}

For $\gamma_1-\gamma_2\approx\beta_1-\beta_2$, the relative weight between the two terms on the right-hand side of $\Delta\bar\lambda$ Eq.~(\ref{Difference_lambda}) is controlled by the ratio $|\phi(R)|/v^2$. Thus, in the case $v\gg\sqrt{|\phi(R)|}$, the probability $P(\kappa^\perp)$ can be approximated by
\begin{equation}
\frac{\Delta E}{\hbar}\Delta\bar\lambda\approx\frac{\phi(R)}{\hbar c^4}\Delta E Lv (\gamma_1-\gamma_2).
\end{equation}
Otherwise, for $v\ll\sqrt{|\phi(R)|}$ we have
\begin{equation}
\frac{\Delta E}{\hbar}\Delta\bar\lambda\approx-\frac{\phi(R)}{\hbar c^4}\Delta E L\frac{\phi(R)}{v} (\beta_1-\beta_2).
\end{equation}
Thus, depending on the speed value $v$, the probability $P(\kappa^\perp)$ captures the non-orthogonality generated by a difference in the values of $\beta$ or $\gamma$ post-Newtonian parameters. In what follows, we consider a quantum system on a trajectory in a constant gravitational field equivalent to that generated by the Earth at a radius $R=7\times10^{6}$ m and thus $\sqrt{|\phi(R)|}=7\times10^3$ m/s.

Our scheme assumes that the quantum clock state remains coherent as the clock propagates. However, quantum states have a natural lifetime $\Delta\tau_{lf}$, after which coherence is lost. Thereby, the time the quantum clock propagates must be shorter than the natural lifetime, or equivalently, for a certain speed of the clock the propagation length $L$ cannot exceed a certain maximum value $L_{\rm{max}}$, that is,
\begin{equation}
    L\le L_{\rm{max}}=\frac{\Delta\tau_{lf}}{\lambda}v.
\end{equation}

Our first choice for quantum clocks is that of freely propagating atoms. In this case, we have an energy difference $\Delta E=6.62\times10^{-19}$ J and a lifetime $\Delta\tau_{lf}=10^{-7}$ s at most. For metric theories such as $\gamma_1-\gamma_2=\beta_1-\beta_2\approx 1$, this choice leads to probabilities $P(\kappa^\perp)$ on the order of $10^{-6}$ for values of $L$ and $v$ respecting the constraints $L\le L_{max}$ and $v\ll c$.  Other choices for a quantum clock are molecules, neutrons, and electrons. However, these also lead to small discrimination probabilities. This discouraging result can be greatly improved by using atomic nuclei as quantum clocks. Recently, the transition from the ground state to the nuclear isomer state of Th-229 was experimentally observed at $\Delta E_{\rm Th}=1.33873708\times10^{-18}$ J. This was achieved by resonant excitation of the transition with the aid of a benchtop tunable vacuum UV laser system. In this experiment, Th-229-doped $\rm{CaF_2}$ crystals were used and a radiative lifetime $\Delta\tau_{\rm{lf}}=630$ s was observed. This was estimated to correspond to $\Delta\tau_{\rm{lf}}=1740$ s for a Th-229 nucleus in vacuum. \cite{Tiedau2024}. Internal conversion is another decay channel where the nuclear transition energy is transferred to a bound electron in the atom, which is then ionized. This process occurs on a time scale on the order of $10^{-5}$ s \cite{Karpeshin2007,Tkalya2015}. For this reason, we consider a bare Th-229 nucleus. This leads to a relative change in nuclear frequency between the neutral Th-229 and its bare nucleus by 1\% \cite{Dzuba2023}. The possibility of driving the ground state of the nucleus of Th-229 to the first excited state by the resonant interaction with a laser makes possible the measurement of the observable $O$ Eq.~(\ref{ObservableO}). This is done by driving the nucleus through a unitary evolution $U^\dagger$, such that $U|0\rangle=|\psi_1\rangle$ and $U|1\rangle=|\psi_1^\perp\rangle$, and observing the decay of the excited state by fluorescence.

\begin{figure}[t!]
    \centering
    \includegraphics[width=0.45\textwidth]{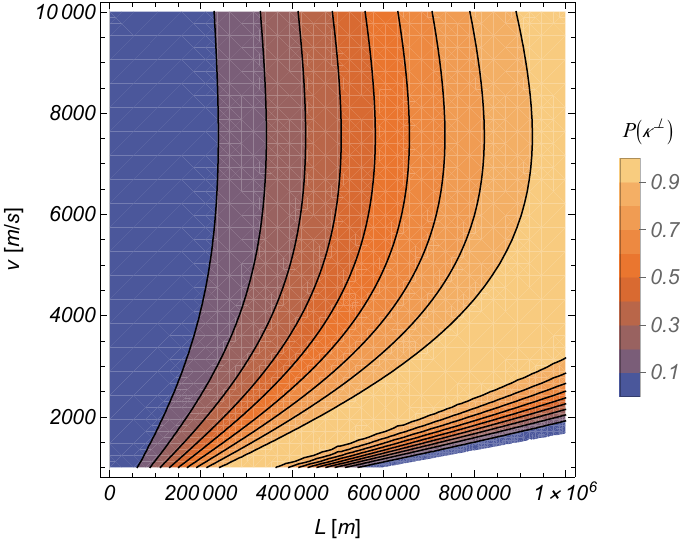}
    \caption{Success probability $P(\kappa^{\perp})$ Eq.~\eqref{Prob_delta_first_proposal} as a function of $L$ and $v$ for a Th-229 nucleus with $\Delta E_{\rm Th}$, $\Delta\tau_{lf}=600$ s, a constant gravitational field equivalent to that generated by the Earth at a radius $R=7\times10^{6}$ m, and $\gamma_2-\gamma_1=\beta_2-\beta_1=2$.}
\label{Fig2}
\end{figure}

\begin{figure}[t!]
    \centering
    \includegraphics[width=0.45\textwidth]{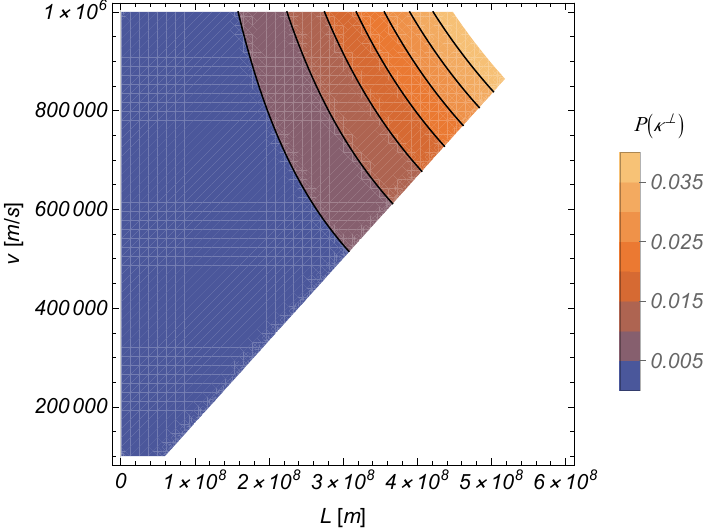}
    \caption{Success probability $P(\kappa^{\perp})$ Eq.~\eqref{Prob_delta_first_proposal} as a function of $L$ and $v$ for a Th-229 nucleus with $\Delta E_{\rm Th}$ J, $\Delta\tau_{\rm{lf}}=600$ s, a constant gravitational field equivalent to that generated by the Earth at a radius $R=7\times10^{6}$ m, and $\gamma_2-\gamma_1=\beta_2-\beta_1=10^{-5}$.}
\label{Fig3}
\end{figure}

Figure~\ref{Fig2} shows the level curves for the probability $P(\kappa^\perp)$ for the thorium nuclei as a function of the propagation length $L$ and the speed $v$ for $\gamma_1-\gamma_2=\beta_1-\beta_2=2$. As is apparent from this figure, there is a large parameter region where the probability $P(\kappa^\perp)$ reaches values above 0.9. In particular, for $L\approx400$ km and $v\approx 2$ km/s the probability is already close to one. Furthermore, for large values of $L$, it is possible to scan a wide range of velocity values with high probability values, so that it is possible to test the origin of nonorthogonality. Thus, a quantum clock consisting of Th-229 nuclei makes it possible to find values of the propagation length and speed such that $P(\kappa^\perp)$ does not vanish, allowing us to rule out a given metric theory.  The length scale can be reduced by increasing the speed of the quantum clock. For example, for $L=3.2$ km and $v=300$ km/s, the success probability is approximately $7\times10^{-2}$.

A more complex situation arises when the values of $\beta_1-\beta_2$ and $\gamma_1-\gamma_2$ are minimal. This occurs when the metric theories to be discriminated have very close post-Newtonian parameters, for example, when trying to rule out a metric theory whose parameters are within the uncertainty of a metric that we suspect describes spacetime. This is shown in Fig.~\ref{Fig3}, where our aim is to distinguish a metric theory with $\gamma_1=\beta_1=1$ from a theory with $\gamma_2=\beta_2=1+10^{-5}$, which leads to probabilities $P(\kappa^\perp)$ of the order of $10^{-2}$ when choosing Th-229 as the quantum clock. The current uncertainty in the post-Newtonian parameters for General Relativity is in the order of $10^{-5}$ \cite{Wei2022}. We see from Fig.~\ref{Fig3} that the parameter region is severely reduced due to constraints $L\le L_{\rm{max}}$ and $v\ll c$, where the highest values of $P(\kappa^\perp)$ are observed for $L$ and $v$ large. 

The possibility of falsifying a given metric theory can be implemented by measuring the observable $O$ Eq.~(\ref{ObservableO}). However, having a sufficiently large probability of the appropriate measurement outcome depends on a careful choice of the energy difference $\Delta E$, gravitational potential $\phi(R)$ and the velocity $v$, for which it is possible to find a length $L$ such that $P(\kappa^\perp)$ does not vanish. In addition, the choice of $L$ depends on the difference $\beta_1-\beta_2$ and $\gamma_1-\gamma_2$, which is in principle unknown. We might, however, suspect that a given metric theory describes spacetime, in which case we can find an optimal length $L_{\rm opt}$ given by
\begin{eqnarray}
    L_{\rm opt} &=& \frac{2 \pi c^4 v \hbar}{\Delta E \phi(R) ( (\beta_1-\beta_2) \phi(R) + (\gamma_1-\gamma_2) v^2)}.
    \label{L_min_Clock_states_First_proposal}
\end{eqnarray}
Placing the second observer at $L_{\rm opt}$ ensures that we can observe a deviation from the metric theory with $\gamma=\gamma_1$ and $\beta=\beta_1$ with the highest probability.

 In our previous analyses, we have assumed that measuring a single quantum clock is sufficient to refute a given hypothesis. However, the associated success probability may be relatively low. The use of an ensemble of quantum clocks, prepared identically and independently, can increase the success probability. Let us now assume that we have at our disposal an ensemble of $N$ quantum clocks, each initially prepared in the state $|\tau\rangle$. After propagation, each quantum clock is described by the state $|\psi\rangle$ Eq.~(\ref{quantumclockstate}). Thus, the collective state of the quantum clock ensemble is $|\psi\rangle^{\otimes N}$. On each quantum clock, we measure the observable $O$, see Eq.~(\ref{ObservableO}), which leads to a string of size $N$ formed by the eigenvalues detected on each quantum clock. The only chain that does not allow us to refute the hypothesis is the one formed by $N$ eigenvalues $\kappa$. Consequently, the success probability $P_S^{(N)}$ is given by
\begin{equation} 
P_S^{(N)}=1-P(\kappa)^N=1-\cos^{2N}\left(\frac{\Delta E}{2\hbar}\Delta\bar\lambda\right).
\label{Ps_N_Clocks}
\end{equation}
In the particular case where $\Delta E\Delta\bar\lambda/\hbar$ is small, the success probability can be approximated as 
\begin{equation}
P_S^{(N)}\approx1-e^{-N(\frac{\Delta E}{2\hbar}\Delta\bar\lambda)^2}, 
\label{PSNapprox}
\end{equation}
up to leading order.  This expression shows that the probability of success exponentially approaches 1 as the size of the ensemble $N$ increases. This suggests that the size of the ensemble acts as a multiplier, effectively amplifying the value of $\Delta E\Delta\bar\lambda/\hbar$ by a factor of $\sqrt{N}$. In the case $v\gg\sqrt{|\phi(R)|}$, we have
\begin{equation}
\sqrt{N}\frac{\Delta E}{2\hbar}\Delta\bar\lambda\approx \frac{\phi(R)}{\hbar c^4}\Delta E Lv (\gamma_1-\gamma_2)\sqrt{N}.
\end{equation}
Thus, small values of $\gamma_1-\gamma_2$ can be balanced by a large enough value of $\sqrt{N}$. For example, for $\gamma_1-\gamma_2\approx 10^{-5}$, we can choose $\sqrt{N}\approx 10^{5}$. Consequently, the resulting success probability now resembles the values observed when $\gamma_1-\gamma_2\approx 1$. Large samples of Th-229 have already been reported. For example, Th-229-doped $\rm{CaF_2}$ crystals at a concentration of $10^{18}$ atoms/cm$^3$ \cite{Beeks2023,Tiedau2024}. Thorium-229 production has also been reported \cite{Hogle2016} in reactors from Ra-226 and Ac-227 targets with yields of (74.0$\pm$7.4) MBq/g and (1200$\pm$50) MBq/g, respectively, after approximately 26 days.

\begin{figure}[t!]
    \centering
    \includegraphics[width=0.45\textwidth]{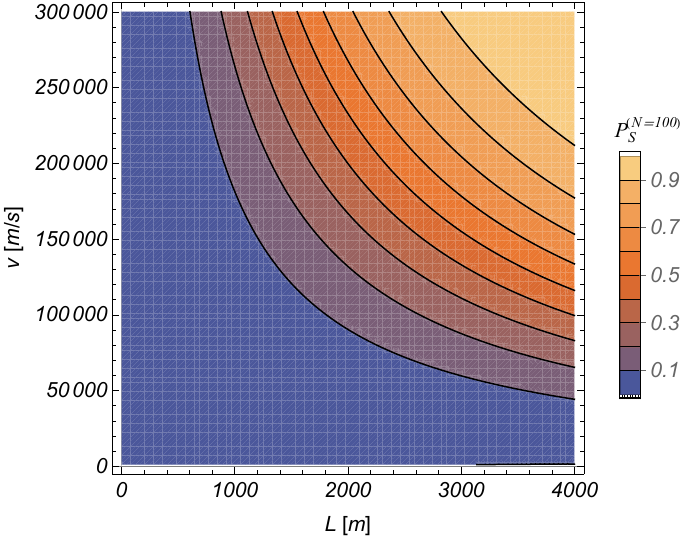}
    \caption{Success probability $P_S^{(N)}$ Eq.~\eqref{Ps_N_Clocks} with $N=100$ as a function of $L$ and $v$ for a Th-229 nucleus with $\Delta E_{\rm Th}$ , $\Delta\tau_{lf}=600$ s, a constant gravitational field equivalent to that generated by the Earth at a radius $R=7\times10^{6}$ m, and $\gamma_2-\gamma_1=\beta_2-\beta_1=2$.}
\label{Fig2_Ps_N_Clocks}
\end{figure}

\begin{figure}[t!]
    \centering
    \includegraphics[width=0.45\textwidth]{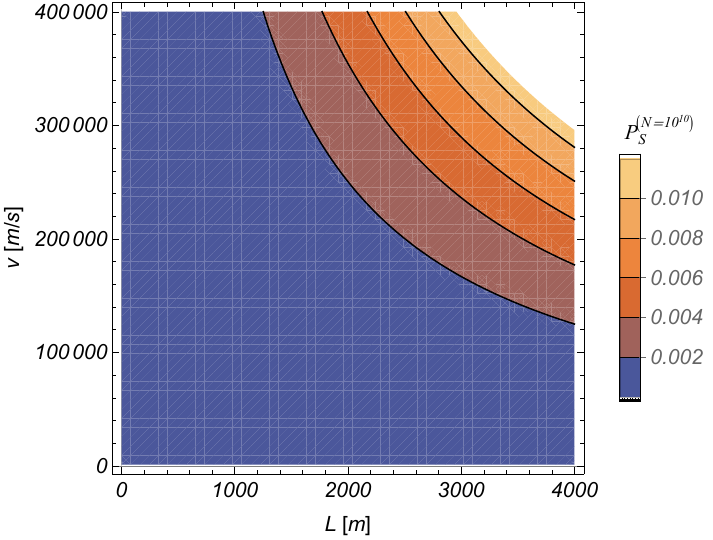}
    \caption{Success probability $P_S^{(N)}$ Eq.~\eqref{Ps_N_Clocks} with $N=10^{10}$ as a function of $L$ and $v$ for a Th-229 nucleus with $\Delta E_{\rm Th}$, $\Delta\tau_{lf}=600$ s, a constant gravitational field equivalent to that generated by the Earth at a radius $R=7\times10^{6}$ m, and $\gamma_2-\gamma_1=\beta_2-\beta_1=10^{-5}$.}
\label{Fig3_Ps_N_Clocks}
\end{figure}

Figures~(\ref{Fig2_Ps_N_Clocks}) and (\ref{Fig3_Ps_N_Clocks}) show the effect of a $N$-quantum clock ensemble on the success probability $P_{S}^{(N)}$ Eq.~(\ref{Ps_N_Clocks}) as a function of $L$ and $v$ for $L$ up to 4 km and $v$ up to 300 km/s for a Th-229 nucleus with $\Delta E_{\rm Th}$, $\Delta\tau_{lf}=600$ s, a constant gravitational field equivalent to that generated by the Earth at a radius $R=7\times10^{6}$ m. Figure~(\ref{Fig2_Ps_N_Clocks}) shows the case $\gamma_2-\gamma_1=\beta_2-\beta_1=2$ and $N=100$ and Fig.~(\ref{Fig3_Ps_N_Clocks}) the case $\gamma_2-\gamma_1=\beta_2-\beta_1=10^{-5}$ and $N=10^{10}$. In both cases, the probability of success $P_{S}^{(N)}$ achieves a large increase for small lengths and speeds compared to case $N=1$.

\subsection{Minimum error discrimination}

In the previous subsection, we have analyzed a simple scheme to rule out or refute a given metric theory within a set of metric theories, which is achieved by measuring the observable $O$ Eq.~(\ref{ObservableO}). This contains in its spectrum the quantum state $\ket{\psi_1^\perp}$, whose detection can only occur if the spacetime is described by a metric theory whose corresponding quantum state is different from $\ket{\psi_1}$. 

A different scenario arises when we attempt to discriminate between two different metric theories, i.e., we have two candidate metric theories to describe spacetime and we need to determine which one is correct. In this case, the measurement to be performed should have the smallest possible probability of misidentification. This is exactly the scenario where minimum error discrimination is used. This discrimination strategy allows us to incorporate a priori information in the form of probabilities that we assign to each metric theory to describe spacetime. The minimal average probability of misidentifying the metric theories is given by the Helstrom bound Eq.~(\ref{HelstromBoundpurestates}), which in our case becomes
\begin{equation}
P_H=\frac{1}{2}\left(1-\sqrt{1-2\eta_1\eta_2(1+\cos(\frac{\Delta E}{\hbar}\Delta \bar\lambda))}\right),
\end{equation}
where $\eta_1$ and $\eta_2=1-\eta_1$ are the a priori probabilities assigned to each metric theory. Assuming equal a priori probabilities, the Helstrom bound becomes

\begin{equation}
P_H=\frac{1}{2}(1-|\sin(\frac{\Delta E}{2\hbar}\Delta\bar\lambda)|).
\label{ME-Error-Probability}
\end{equation}

We can compare this strategy with the simple approach based on $O$ Eq.~(\ref{ObservableO}). According to this, the detection of the eigenvalue $\kappa^\perp$ ($\kappa$) is interpreted as detecting the state $\ket{\psi_2}$ ($\ket{\psi_1}$). Thus, the average error probability for identifying the state is
\begin{equation}
P_E=\frac{1}{2}-\frac{1}{2}P(\kappa^\perp).   
\end{equation}
This can be compared with $P_H=1/2-\sqrt{P(\kappa^\perp)}/2$, which shows that the minimum error discrimination strategy provides a higher average success probability given by 

\begin{equation}
P_{S,\rm{min}}=\frac{1}{2}(1+|\sin(\frac{\Delta E}{2\hbar}\Delta\bar\lambda)|),
\label{ME-Succes-Probability}
\end{equation}
for all values of $\Delta E$ and $\Delta\bar\lambda$.

\begin{figure}[t!] 
    \centering
    \includegraphics[width=0.45\textwidth]{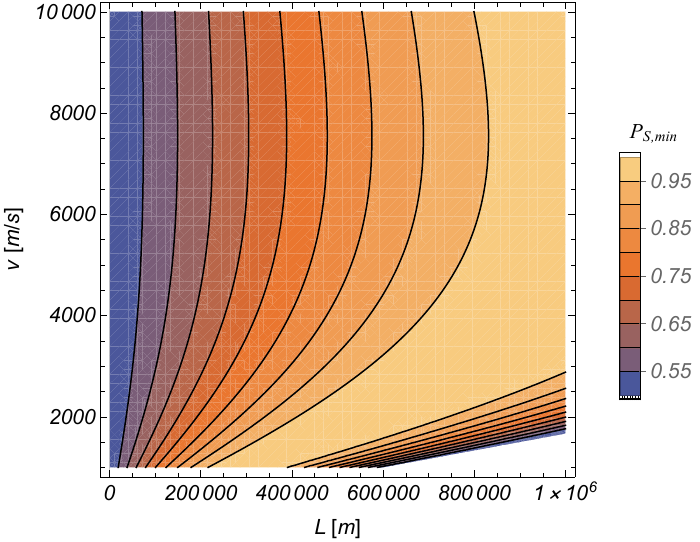}
    \caption{Success probability $P_{S, \rm{min}}$ Eq.~\eqref{ME-Succes-Probability} as a function of $L$ and $v$ for a Th-229 nucleus with $\Delta E_{\rm Th}$, $\Delta\tau_{lf}=600$ s, a constant gravitational field equivalent to that generated by the Earth at a radius $R=7\times10^{6}$ m, and $\gamma_2-\gamma_1=\beta_2-\beta_1=2$.}
    \label{Fig4}
\end{figure}

The average success probability $P_{S,\rm{min}}$ Eq.~(\ref{ME-Succes-Probability}) provided by the minimum error discrimination strategy is shown in Fig.~\ref{Fig4} as a function of length $L$ and speed $v$ for the case of a Th-229 nucleus. We discriminate a metric theory with $\gamma_1=\beta_1=1$ from a theory with $\gamma_2=\beta_2=3$. As this figure shows, the discimination between two competing theories has a higher probability than ruling out a given theory.  The length scale can be reduced by increasing the speed of the quantum clock; for example, for $L=3.2$ km and $v=300$ km/s, the success probability is approximately $0.54$.

\begin{figure}[t!]
    \centering
    \includegraphics[width=0.48\textwidth]{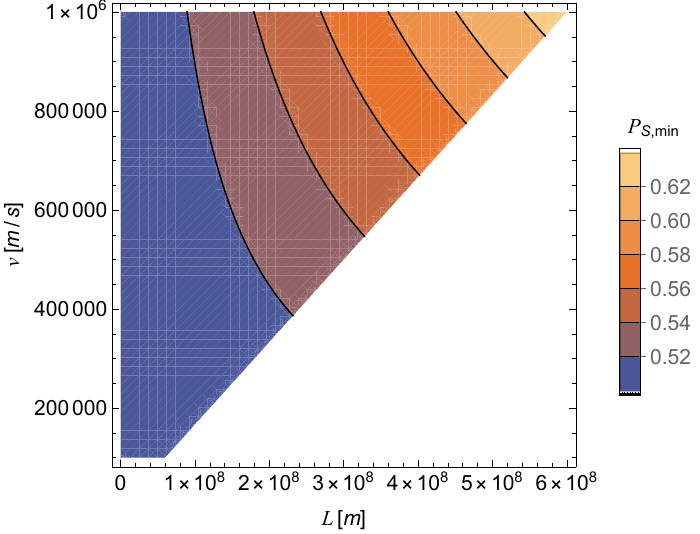}
    \caption{Success probability $P_{S,\rm{min}}$ Eq.~\eqref{ME-Succes-Probability} as a function of $L$ and $v$ for a Th-229 nucleus with $\Delta E_{\rm Th}$, $\Delta\tau_{\rm{lf}}=600$ s, a constant gravitational field equivalent to that generated by the Earth at a radius $R=7\times10^{6}$ m, and $\gamma_2-\gamma_1 = \beta_2-\beta_1 = 10^{-5}$.}
    \label{Fig5}
\end{figure}

It is possible to distinguish between metric theories, even deterministically. This occurs whenever it is possible to choose a length such that the states $\ket{\psi_1}$ and $\ket{\psi_2}$ become orthogonal. This length is given by
\begin{equation}
L_{S,\rm opt} = \frac{\pi c^4 v \hbar}{\Delta E \phi(R) \left((\beta_1-\beta_2)  \phi(R) + (\gamma_1-\gamma_2 ) v^2\right)}.
\end{equation}
An interesting result of using minimum error discrimination is that, as Fig.~\ref{Fig5} indicates, the probability of discriminating very close metric theories becomes higher than in the case of the simple discrimination scheme. In particular, the success probability is one order of magnitude higher. 

 As in the case of the simple discrimination scheme, an improvement in the probability of success can be obtained by considering a large sample of equally prepared quantum clocks. In the symmetric case $\eta_1=\eta_2$, the error probabilities $P(1||\psi_2\rangle\langle\psi_2|)$ and $P(2||\psi_1\rangle\langle\psi_1|)$ are equal, and thus they are equal to $P_H$, see Eq.~(\ref{ME-Error-Probability}). Thus, a sample of $N$ quantum clocks, prepared equally and independently, will be described by the states $|\psi_1\rangle^{\otimes N}$ and $|\psi_2\rangle^{\otimes N}$, each with probability $1/2$. If each quantum clock undergoes the minimum error discrimination strategy, the total success probability is
\begin{equation}
P^{(N)}_{S}=1-P_H^N,
\end{equation}
or equivalently

\begin{equation}
P^{(N)}_{S,\rm{min}}=1-\frac{1}{2^N}(1-|\sin(\frac{\Delta E}{2\hbar}\Delta\bar\lambda)|)^N.
\label{ME-Succes-Probability-NClocks}
\end{equation}
In the particular case where $\Delta E\Delta\bar\lambda/\hbar$ is small, this success probability can be approximated as
\begin{equation}
P^{(N)}_{S,\rm{min}}\approx1-\frac{1}{2^N}e^{-N(\frac{\Delta E}{2\hbar}\Delta\bar\lambda+\frac{1}{2}(\frac{\Delta E}{2\hbar}\Delta\bar\lambda)^2)},  
\label{MEsepN}
\end{equation}
up to leading order. This expression aproaches 1 faster than $P_S^{(N)}$ in Eq.~\eqref{PSNapprox}.

It is possible to increase the success probability even further by resorting to a collective measurement. In this case, the measurement acts on the $N$ quantum clocks simultaneously and has only two outcomes. The success probability is thus given by
\begin{equation}
P^{(N)}_{S,\rm{min}}=1-\frac{1}{2}\left(1-\sqrt{1-|\langle\psi_0|\psi_1\rangle|^{2N}}\right),   
\end{equation}
or equivalently
\begin{equation}
P^{(N)}_{S,\rm{min}}=\frac{1}{2}+\frac{1}{2}\sqrt{1-\cos^{2N}({\frac{\Delta E}{2\hbar}\Delta\bar\lambda})}.
\end{equation}
Implementing a collective measurement traditionally requires the application of a unitary operator that entangles all members of the ensemble. In the case of quantum clocks, gravity itself has been considered to be a source of entanglement between them \cite{Castro2017}. However, implementing collective measurements on the atomic nuclei remains a challenge.

\begin{figure}[h!] 
    \centering
    \includegraphics[width=0.45\textwidth]{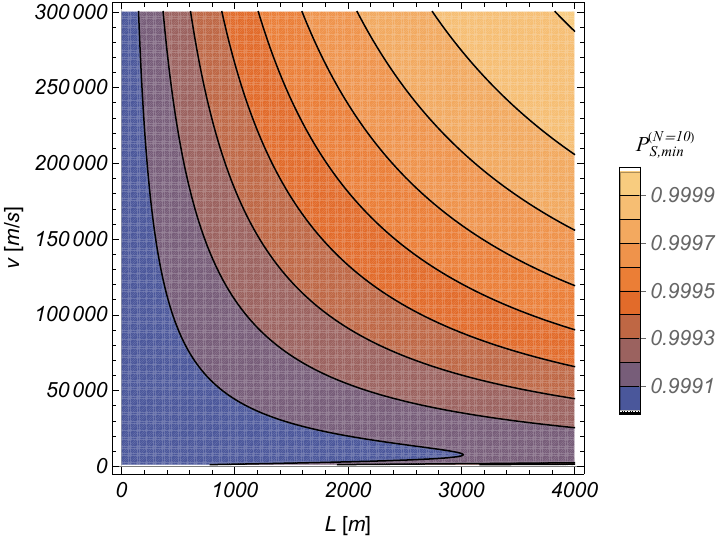}
    \caption{Success probability $P_{\rm{S,min}}^{(N)}$ Eq.~\eqref{ME-Succes-Probability-NClocks} with $N=10$ as a function of $L$ and $v$ for a Th-229 nucleus with $\Delta E_{\rm Th}$, $\Delta\tau_{\rm{lf}}=600$ s, a constant gravitational field equivalent to that generated by the Earth at a radius $R=7\times10^{6}$ m, and $\gamma_2-\gamma_1=\beta_2-\beta_1=2$.}
    \label{Fig7}
\end{figure}

\begin{figure}[h!]
        \centering
        \includegraphics[width=0.48\textwidth]{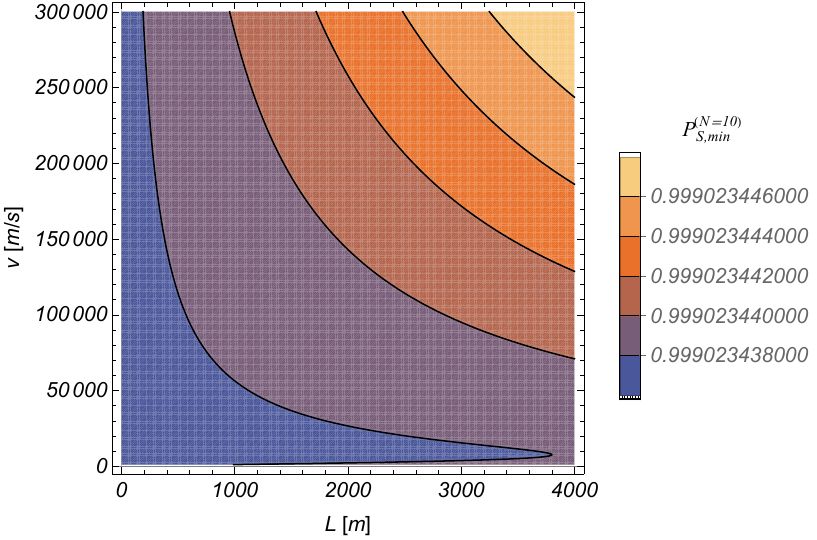}
        \caption{Success probability $P_{\rm{S,min}}^{(N)}$ Eq.~\eqref{ME-Succes-Probability-NClocks} with $N=10$ as a function of $L$ and $v$ for a Th-229 nucleus with $\Delta E_{\rm Th}$, $\Delta\tau_{\rm{lf}}=600$ s, a constant gravitational field equivalent to that generated by the Earth at a radius $R=7\times10^{6}$ m, and $\gamma_2-\gamma_1 = \beta_2-\beta_1 = 10^{-5}$.}
        \label{Fig8}
\end{figure}

The behavior of $P_{\rm{S,min}}^{(N)}$ as a function of $L$ and $v$, according to Eq.~(\ref{ME-Succes-Probability-NClocks}), is displayed in Figs.~(\ref{Fig7}) and (\ref{Fig8}) for $\gamma_2-\gamma_1=\beta_2-\beta_1=2$ and $\gamma_2-\gamma_1=\beta_2-\beta_1=10^{-5}$, respectively. We consider $L$ up to 4 km and $v$ up to 300 km/s for a Th-229 nucleus with $\Delta E=1.33873708\times10^{-18}$ J, $\Delta\tau_{lf}=600$ (s), a constant gravitational field equivalent to that generated by the Earth at a radius $R=7\times10^{6}$ (m). In both cases, the set size is $N=10$, achieving a probability of success close to 0.9999, also in both cases.

\subsection{Unambiguous discrimination}

So far, we have studied two strategies for discriminating metric theories. The first and simple approach allows us to conclusively rule out or refute a given metric theory with respect to a set of possible metric theories. The second approach, based on minimum error discrimination, allows us to distinguish between any two metric theories, albeit with a minimum misidentification error. However, there exists a length $L_O$ such that this error vanishes, making it possible for an error-free discrimination.  

It is possible to discriminate between two metric theories without error, even if the length value cannot be fixed in $L_O$. This is achieved through unambiguous state discrimination, which allows nonorthogonal states to be detected without error at the expense of introducing an inconclusive event. This does not allow conclusions to be drawn about the states that are discriminated. Following Eq.~(\ref{ProbabilityUD}), the optimal average success probability for equal a priori probabilities is given by

\begin{equation}
P_{S,\rm{ud}}=1-|\cos(\frac{\Delta E}{2\hbar}\Delta\bar\lambda)|.  
\label{UD-Probability}
\end{equation}
This turns out to be less than or equal to $P_S$ Eq.~(\ref{ME-Succes-Probability}), but has the advantage that the measurement results associated with the states do not lead to misidentifications. Thus, a single detection allows us to determine which metric theory describes spacetime. 

\begin{figure}[t!]
    \centering
    \includegraphics[width=0.45\textwidth]{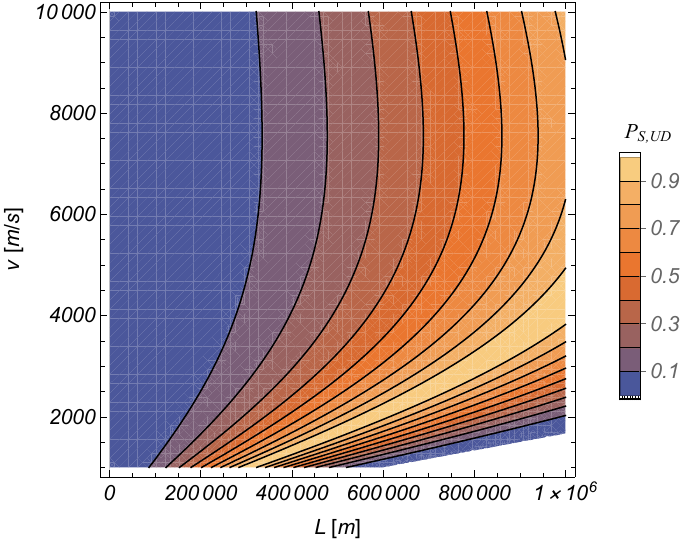}
    \caption{Success probability $P_{S,\rm{ ud}}$ Eq.~\eqref{UD-Probability} as a function of $L$ and $v$ for a Th-229 nucleus with $\Delta E_{\rm Th}$, $\Delta\tau_{\rm{lf}}=600$ s, a constant gravitational field equivalent to that generated by the Earth at a radius $R=7\times10^{6}$ m, and $\gamma_2-\gamma_1=\beta_2-\beta_1=2$.}
    \label{Fig6_UD-Probability}
\end{figure}

\begin{figure}[t!]
    \centering
    \includegraphics[width=0.48\textwidth]{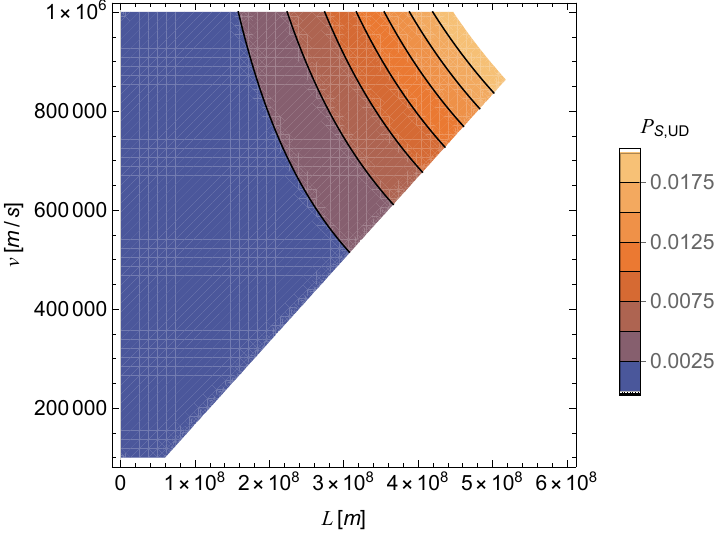}
    \caption{Probability of detection $P_{S,\rm{ud}}$ Eq.~\eqref{UD-Probability}as a function of $L$ and $v$ for a Th-229 nucleus with $\Delta E_{\rm Th}$, $\Delta\tau_{\rm{lf}}=600$ s, a constant gravitational field equivalent to that generated by the Earth at a radius $R=7\times10^{6}$ m, and $\gamma_2-\gamma_1 = \beta_2-\beta_1 = 10^{-5}$.}
    \label{Fig7_UD-Probability}
\end{figure}

The behavior of $P_{\rm UD}$ is shown in Fig.~\ref{Fig6_UD-Probability} as a function of length $L$ and speed $v$ for the Th-229 nucleus. As this figure shows, there is a clear decrease in the value of the probability of success $P_{\rm US}$ compared to the previous schemes, although it remains in the same order of magnitude. Fig.~\ref{Fig7_UD-Probability} shows $P_{\rm UD}$ in the discrimination of two close metric theories. In this case, there is a clear decrease of $P_{\rm UD}$ of approximately one order of magnitude with respect to the use of minimum error, which makes it comparable to the results obtained by the simple discrimnation scheme. However, we benefit from the key feature of unambiguous discrimination: a single detection allows us to conclusively determine the metric theory that describes space-time.

 The probability of unambiguously distinguishing between metric theories can also be enhanced by using an identically and independently prepared ensemble of quantum clocks. If each member of the ensemble undergoes the unambiguous discrimination strategy, the only instance that does not allow for a conclusive result is when all measurements are inconclusive. Thus, the success probability is given by
\begin{equation}
P_{S,\rm{ud}}^{(N)}=1-|\cos(\frac{\Delta E}{2\hbar}\Delta\bar\lambda)|^N,  
\label{UD-Probability-NClocks}
\end{equation}
which for an argument close to zero can be approximated by
\begin{equation}
P_{S,\rm{ud}}^{(N)}\approx1-e^{-N(\frac{\Delta E}{2\sqrt{2}\hbar}\Delta\bar\lambda)^2},
\end{equation}
up to leading order. This expression shows an exponential increase in the probability of success towards 1, although at a reduced rate compared to the probability of success Eq.~(\ref{MEsepN}) obtained by minimum error discrimination with separable measurements.

\begin{figure}[h!]
    \centering
    \includegraphics[width=0.45\textwidth]{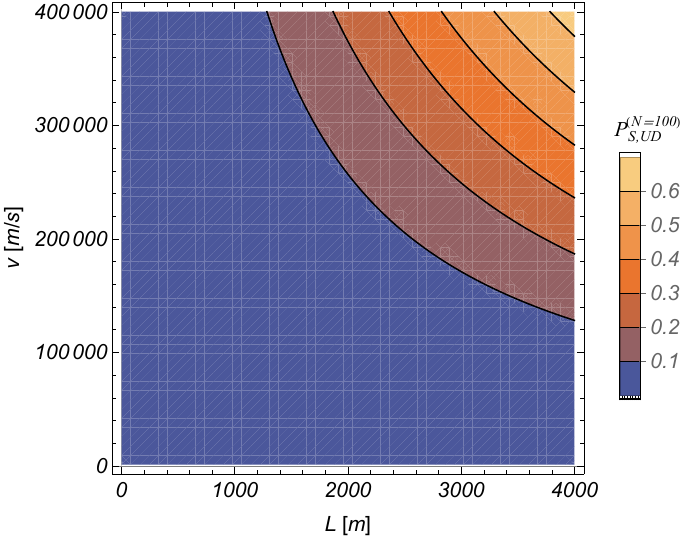}
    \caption{Success probability $P_{S,\rm{ud}}^{(N)}$ Eq.~\eqref{UD-Probability-NClocks} with $N=10$ as a function of $L$ and $v$ for a Th-229 nucleus with $\Delta E_{\rm Th}$, $\Delta\tau_{lf}=600$ s, a constant gravitational field equivalent to that generated by the Earth at a radius $R=7\times10^{6}$ m, and $\gamma_2-\gamma_1=\beta_2-\beta_1=2$.}
    \label{Fig11}
\end{figure}

\begin{figure}[h!]
    \centering
    \includegraphics[width=0.48\textwidth]{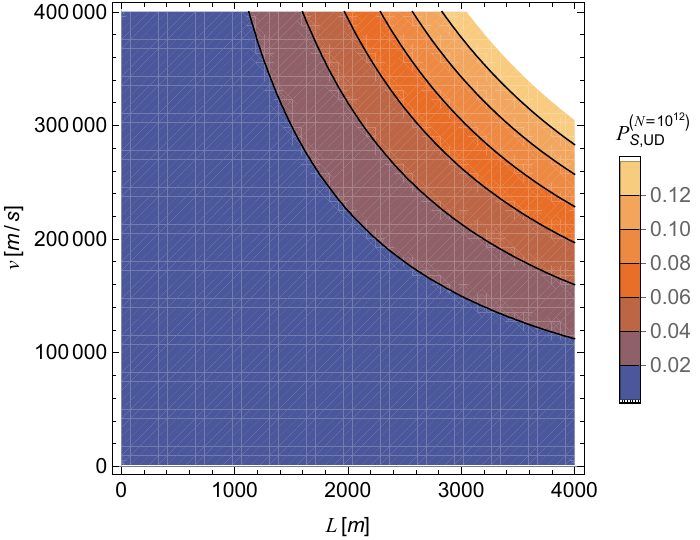}
    \caption{Probability of detection $P_{S,\rm{ud}}^{(N)}$ Eq.~\eqref{UD-Probability-NClocks} with $N=10^{12}$ as a function of $L$ and $v$ for a Th-229 nucleus with $\Delta E_{\rm Th}$, $\Delta\tau_{\rm{lf}}=600$ s, a constant gravitational field equivalent to that generated by the Earth at a radius $R=7\times10^{6}$ m, and $\gamma_2-\gamma_1 = \beta_2-\beta_1 = 10^{-5}$.}
    \label{Fig12}
\end{figure}

Figures~(\ref{Fig11}) and (\ref{Fig12}) show the success probability $P_{S,\rm{ud}}^{(N)}$ as a function of $L$ and $v$ for a Th-229 nucleus with $\Delta E=1.33873708\times10^{-18}$ J, $\Delta\tau_{\rm{lf}}=600$ (s), a constant gravitational field equivalent to that generated by the Earth at a radius $R=7\times10^{6}$ (m). Fig.~(\ref{Fig11}) shows the case $\gamma_2-\gamma_1=\beta_2-\beta_1=2$ and $N=100$ and Fig.~(\ref{Fig12}) the case $\gamma_2-\gamma_1=\beta_2-\beta_1=10^{-5}$ and $N=10^{12}$.
As these figures show, the probability of success $P_{S,\rm{ud}}^{(N)}$ achieves a large increase for small lengths and speeds compared to case $N=1$.

\section{Discussion and Conclusion}
\label{SECTION4}

We have studied the problem of discriminating between metric theories. We have shown that, within the post-Newtonian approach, the quantum state of a massive two-dimensional quantum clock becomes, after propagating at a low speed and in a weak gravitational field, a function of the parameters $\gamma$ and $\beta$. Thus, the discrimination between metric theories becomes a discrimination between quantum states of the quantum clock. Since the number of metric theories is greater than two, the problem of discrimination becomes difficult. We have applied to this problem three different quantum state discrimination strategies. We first resorted to measuring a rank-two observable. This approach allows ruling out or refuting a particular metric theory with respect to a set of metric theories, although it is not possible to determine which theory within the set describes spacetime. A single detection event enough for this purpose, albeit its probability can become very small. A second approach, minimum-error discrimination, leads to an increase in the average probability of a successful discrimination when dealing with only two metric theories. However, this occurs at the expense of accepting a small misidentification error. This can be improved with the help of unambiguous state discrimination, which identifies different metric theories without error but with a lower probability of success. In the first and third approach, a single detection event is enough to discard a given metric theory. 

The probability of success of the three discrimination strategies is a harmonic function of the difference in proper time corresponding to each state of the quantum clock multiplied by the propagation length and the energy difference between the eigenstates of the quantum clock. This favors the use of an atomic nucleus, such as Th-229, instead of atoms, molecules, neutrons, and electrons, in which cases the probability of success becomes vanishingly small.

A Th-229 nucleus allows us to discriminate between metric theories such that $\Delta\beta=\Delta\gamma\approx1$ with propagation lengths on the order of $10^3$ km and speeds on the order of $10$ km/s, in which case success probabilities close to 1 can be reached in all discrimination schemes studied.  The use of shorter lengths and higher speeds leads to small probabilities. For example, for $L=3.2$ km and $v=300$ km/s, the success probability is of the order of $10^{-2}$. The discrimination of metric theories that have very close post-Newtonian parameters requires lengths on the order of $10^5$ km and speeds on the order of $10^3$ km/s to reach success probabilities on the order of $10^{-2}$. Higher probabilities can be achieved on larger length or speed scales.  Additionally, we have shown that it is possible to use short lengths and low speeds using an ensemble of quantum clocks, each independently prepared in the same initial state. The ensemble of quantum clocks is measured, independently or collectively, with one of the discrimination strategies. This approach leads to a success probability that exponentially approaches 1 as the number of quantum clocks increases. A large enhancement of the success probability is achieved in the case of minimum-error discrimination, where for the case $\gamma_2-\gamma_1=\beta_2-\beta_1=10^{-5}$ an ensemble of 10 quantum clocks and separable measurements lead to a success probability close to 0.999.

Several extensions of our current results are possible. We concentrated on the two post-Newtonian parameters $\gamma$ and $\beta$, which do not fully characterize all possible metric theories. Thus, it would be interesting to extend our results to the case of a larger number of post-Newtonian parameters. However, there are some challenges along this route. A larger number of parameters probably requires a quantum clock of higher dimensions \cite{borregaard2024testingquantumtheorycurved}, which in turn might lead to a more complicated set of states that must be discriminated. Also, we have studied the case of discriminating between two quantum clock states. It is possible to discriminate among more states, but in this case general analytical results for the success probability and the optimal measurement are lacking, leading to the use of numerical techniques for solving the quantum state discrimination problem. Finally, note that other discrimination strategies can be used, such as maximum confidence \cite{Croke2006,Jimenez2011} and fixed rate of inconclusive outcome \cite{ZHANG199925,Gomez2022}.

\section*{Acknowledgements}
OJ was supported by the Universidad Mayor PUENTE-2024-17. AD and FJL were supported by the National Agency of Research and Development (ANID) -- Millennium Science Initiative Program -- ICN17$_-$012. MRT acknowledges the support of Fondecyt postdoctoral project No. $3230407$.

\appendix

\section{Phase-shift}\label{Appendix_Phase_shift}

We consider the Hamiltonian presented in Eq.\thinspace\eqref{Modified_Hamiltonian_PPN} and seek a solution to the corresponding modified Schr{\"o}dinger Equation. We adopt the ansatz $\vert\psi \rangle = e^{i\xi} \vert \psi_{0}\rangle$, where $\vert \psi_{0}\rangle$ is the solution to the free non-relativistic Schr{\"o}dinger Equation governed by the Hamiltonian $H_0 = \hat{p}^2/(2m)$, with $\hat{p}$ being the momentum operator, and $\xi = \xi(t)$.

We now evaluate the left-hand side (LHS) of the Schr{\"o}dinger Equation, $i \hbar \partial \vert \psi \rangle/\partial t$. Substituting the ansatz yields:
\begin{align*}
i\hbar \frac{\partial}{\partial t} \left( e^{i\xi} \vert\psi_0\rangle \right)
&= i\hbar \left( i\dot{\xi} e^{i\xi} \vert\psi_0\rangle + e^{i\xi} \frac{\partial \vert\psi_0\rangle}{\partial t} \right) \\
&= i\hbar \left\{ i\dot{\xi} e^{i\xi} \vert\psi_0\rangle + e^{i\xi} \frac{H_{0}}{i\hbar} \vert\psi_0\rangle \right\}.
\end{align*}
Similarly, we evaluate the right-hand side (RHS) of the Schr{\"o}dinger Equation:
\begin{align}
    H|\psi\rangle = e^{i\xi} H_{0} |\psi_0\rangle + e^{i\xi} \sum_k H^{\mathrm{Rel},k} |\psi_0\rangle.
\end{align}
Equating this result with the LHS and canceling the common terms (including the phase factor $e^{i\xi}$) yields:
\begin{align*}
-\hbar\dot{\xi} |\psi_0\rangle = \sum_k H^{\mathrm{Rel},k} |\psi_0\rangle.
\end{align*}
Finally, we take the expectation value of the preceding equation with respect to $\vert \psi_0\rangle$ and integrate with respect to the time to solve for $\xi$. The first step yields the following:
\begin{align*}
-\hbar\dot{\xi} &= \sum_k \langle H^{\mathrm{Rel},k} \rangle.
\end{align*}
Integrating this expression from $t_0$ to $t$ gives the total phase shift:
\begin{align*}
\Delta \xi &= -\frac{1}{\hbar}\sum_k \int_{t_0}^{t} dt \langle H^{\mathrm{Rel},k} \rangle.
\end{align*}
This result is consistent with the findings in \cite{Zych2011, PhysRevD.55.1964, Rivera-Tapia_2022}. Furthermore, using the definition of the total mass $M$ from Eq.\thinspace\eqref{Mass-Energy_Quantum}, the Hamiltonian takes the form shown in Eq.\thinspace\eqref{TotalHamiltonian}, where the terms $H^{\mathrm{rel}}_{\mathrm{cm}}$, $\lambda$ and the internal clock Hamiltonian $H_{\mathrm{clock}}$ are defined in Eqs.\thinspace\eqref{H_cm}, \eqref{Coupling_operator}, and \eqref{H_clock}, respectively. Therefore, the phase shift can be decomposed as
\begin{align}
    \xi &= -\frac{1}{\hbar}\int_{t_0}^{t}dt \langle H^{\mathrm{rel}}_{\mathrm{cm}} \rangle -\frac{1}{\hbar}\int_{t_0}^{t}dt  \langle \lambda \rangle H_{\mathrm{clock}}.
\end{align}
The expectation value is calculated with respect to the states $|\psi_0\rangle$ of the free Schr{\"o}dinger equation. 

The physical interpretation of this result corresponds to the semiclassical approximation. In this limit, the center-of-mass degrees of freedom (position $\vec{x}$ and momentum $\vec{p}$) are treated as classical variables with well-defined values, equivalent to averaging over their quantum wave packet.

Finally, incorporating the full phase shift into our ansatz, the solution to the modified Schr{\"o}dinger Equation is found to be:
\begin{eqnarray}
    |\psi(t)\rangle = e^{ -(i/\hbar)\int dt \langle H^{\mathrm{rel}}_{\mathrm{cm}} \rangle} |\psi_{0}\rangle e^{-(i/\hbar) \int dt \langle \lambda \rangle H_{\mathrm{clock}}}|\tau_0 \rangle.
\end{eqnarray}
Here, we have explicitly included $|\tau_0 \rangle$ as the initial state of the system's internal degrees of freedom.

Having established the expression for the difference in arrival times, $\Delta \bar{\lambda}$, in Eq.~\eqref{Difference_lambda}, we now proceed to evaluate its quantum mechanical expectation value. For this purpose, we model the quantum state of the particle as a Gaussian wave packet of the form
\begin{equation}
    \vert \psi_0\rangle = \left(\frac{1}{2\pi \sigma_t^2}\right)^{1/4} \exp\left[-\frac{x^2}{4 \sigma_t^2} + i \frac{p_0}{\hbar} \left(x- \frac{p_0 t}{2m} \right)\right],
\end{equation}
where the time-dependent standard deviation is given by $\sigma_t = \sigma_0 \sqrt{1+ \left(\hbar t/(2m\sigma_0^2)\right)^2}$. Here, $\sigma_0$ represents the initial width of the wave packet, and the expression for $\sigma_t$ quantifies its natural dispersion over time.

Evaluating the expectation value of the operator associated with the observable in Eq.~\eqref{Difference_lambda} with respect to the state $\vert \psi_0\rangle$ and considering a linear approximation of the gravitational potential $\phi(R+r) = \phi(R)+ g\cdot r$ yields
\begin{align}
\langle \Delta \lambda \rangle =& \frac{\phi(R)L}{c^4 v} \left[ \Delta \gamma \left(\frac{p_0^2}{m^2} + \frac{\sigma_0^2 \hbar^2}{4m^2 \sigma_0^4 + t^2 \hbar^2}\right) \right. \nonumber \\
& \left. + \Delta \beta \left(\phi(R) + \frac{g^2 (4m^2 \sigma_0^4 + t^2 \hbar^2)}{4c^4 m^2 \sigma_0^2 \phi(R)}\right) \right].
\end{align}
Then, integrating over coordinate time, that is, $\langle \Delta \bar{\lambda} \rangle = \int dt \langle \Delta \lambda \rangle$. Therefore, we obtain
\begin{align} \label{eq:Difference-Arrival-times-Gaussian}
    \langle \Delta \bar{\lambda} \rangle =&
 \, t \left[ \Delta \beta \left(\frac{g^{2} \sigma_{0}^{2}}{c^{4}} + \frac{\phi(R)^{2}}{c^{4}}\right) - \frac{p_{0}^{2} \phi(R) \Delta\gamma}{c^{4} m^{2}} \right]  \nonumber \\
& + \frac{g^{2} \hbar^{2} t^{3} \Delta \beta}{12 c^{4} m^{2} \sigma_{0}^{2}} -  \frac{\hbar \phi(R) \Delta\gamma \arctan{\left(\frac{\hbar t}{2 m \sigma_{0}^{2}} \right)}}{2 c^{4} m}.
\end{align}
Following Eq.~\eqref{eq:mean-lambda}, it is useful to decompose this result into its classical and quantum-mechanical contributions, $\langle \Delta \lambda \rangle = \Delta \bar{\lambda} + \Delta \bar{\lambda}_{\rm quant}$. Here, $\Delta \bar{\lambda}$ is the purely classical expression of Eq.~\eqref{Difference_lambda}. The additional term,

\begin{align}
    \Delta \bar{\lambda}_{\rm quant} =& \, \frac{g^{2} \Delta \beta}{c^{4}} \left( \sigma_{0}^{2} t + \frac{\hbar^{2} t^{3}}{12 m^{2} \sigma_{0}^{2}} \right)\nonumber \\
    & -  \frac{\hbar \phi(R) \Delta \gamma \arctan{\left(\frac{\hbar t}{2 m \sigma_{0}^{2}} \right)}}{2 c^{4} m},
\end{align}
constitutes the quantum correction. This term explicitly depends on $\hbar$ and arises from the quantum nature of the wave packet.

\begin{figure}[h!]
    \centering
    \includegraphics[width=0.48\textwidth]{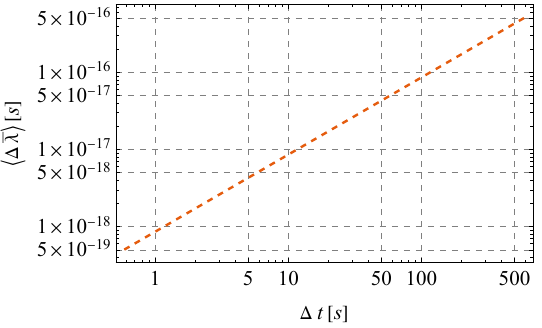}
    \caption{Plot of the expectation value for the arrival time difference, $\langle \Delta \bar{\lambda} \rangle$, as a function of the clock's lifetime, $\Delta t$. The calculation, based on Eq.~\eqref{eq:Difference-Arrival-times-Gaussian}, uses parameters corresponding to a Thorium (Th) clock: velocity $v = 2000\,\mathrm{m/s}$ and initial standard deviation $\sigma_0 = 10^{-3}\,\mathrm{m}$. The gravitational parameters are taken to be $\Delta \gamma = \gamma_2 - \gamma_1 = 2$ and $\Delta\beta = \beta_2 - \beta_1 = 2$.}
    \label{fig:Diff-Expectation-value-lambda-Gaussian}
\end{figure}

\begin{figure}[H]
    \centering
    \includegraphics[width=0.48\textwidth]{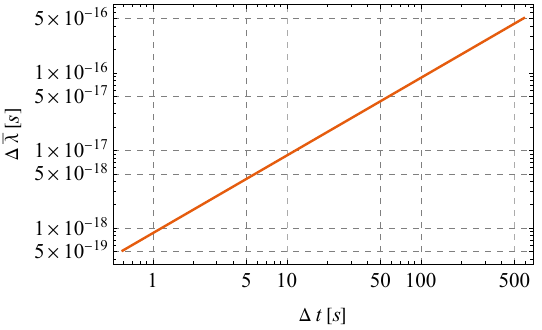}
    \caption{Plot of the expectation value for the arrival time difference, $\langle \Delta \bar{\lambda} \rangle$, as a function of the clock's lifetime, $\Delta t$. The calculation, based on Eq.~\eqref{Difference_lambda}, uses parameters corresponding to a Thorium (Th) clock with velocity $v = 2000\,\mathrm{m/s}$. The gravitational parameters are taken to be $\Delta \gamma = \gamma_2 - \gamma_1 = 2$ and $\Delta\beta = \beta_2 - \beta_1 = 2$.}
    \label{fig:Diff-Expectation-value-lambda-Classical}
\end{figure}

To analyze the behavior of this quantum-corrected arrival time difference and to ascertain the magnitude of the quantum effects, we perform a numerical comparison. We compare the full expectation value of Eq.~\eqref{eq:mean-lambda} with its classical counterpart, Eq.~\eqref{Difference_lambda}. For this analysis, we adopt parameters corresponding to a Thorium atomic clock with a velocity of $v= 2000\,\mathrm{m/s}$ and an initial wave packet width of $\sigma_0 = 10^{-3}\,\mathrm{m}$. Figure~\ref{fig:Diff-Expectation-value-lambda-Gaussian} illustrates the quantum expectation value $\langle \Delta \bar{\lambda} \rangle$ as a function of the clock's lifetime, while Fig.~\ref{fig:Diff-Expectation-value-lambda-Classical} shows the classical result under identical conditions. A direct comparison reveals no discernible difference between the two plots, suggesting that for this parameter regime, the quantum corrections are negligible.

\begin{figure}[H]
    \centering
    \includegraphics[width=0.48\textwidth]{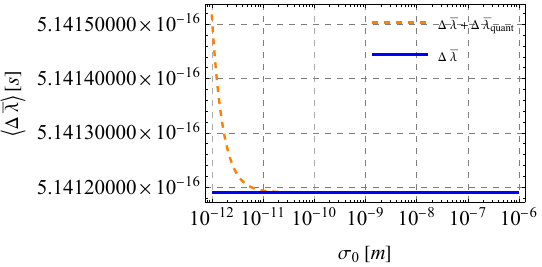}
    \caption{A comparative analysis of the quantum and classical predictions for the arrival time difference as a function of the initial standard deviation, $\sigma_0$. The quantum result, $\langle \Delta \bar{\lambda} \rangle$ (Eq.~\eqref{eq:Difference-Arrival-times-Gaussian}), and the classical result, $\Delta \bar{\lambda}$ (Eq.~\eqref{Difference_lambda}), are plotted for a Thorium (Th) clock. This analysis is performed at a constant clock lifetime of $\Delta t_{\rm lf} = 600\,\mathrm{s}$, with velocity $v= 2000\,\mathrm{m/s}$, and gravitational parameters $\Delta \gamma = 2$ and $\Delta\beta = 2$.}
    \label{fig:Diff-Expectation-value-lambda-Both}
\end{figure}

Furthermore, to investigate the influence of the initial wave packet width, we plot the expectation values for both the quantum and classical cases as a function of the standard deviation $\sigma_0$, with the lifetime fixed at $\Delta t_{\rm lf} = 600\,\mathrm{s}$. The results, depicted in Fig.~\ref{fig:Diff-Expectation-value-lambda-Both}, indicate that the quantum and classical predictions converge for initial standard deviations larger than approximately $\sigma_0 \approx 10^{-11}\,\mathrm{m}$, reinforcing the conclusion that classical dynamics provides an excellent approximation for macroscopic systems.

\bibliography{references}

\end{document}